\documentstyle[twocolumn,prc,aps,epsfig]{revtex}
\input epsf
\newcommand{\be}{\begin{eqnarray}}
\newcommand{\ee}{\end{eqnarray}}

 \newcommand{\gsim}{\mathrel{\hbox{\rlap{\lower.55ex \hbox {$\sim$}}
                   \kern-.3em \raise.4ex \hbox{$>$}}}}
\newcommand{\lsim}{\mathrel{\hbox{\rlap{\lower.55ex \hbox {$\sim$}}
                   \kern-.3em \raise.4ex \hbox{$<$}}}}

\def\roughly#1{\mathrel{\raise.3ex\hbox{$#1$\kern-.75em%
\lower1ex\hbox{$\sim$}}}}
\def\lsim{\roughly<}
\def\gsim{\roughly>}

\def\la{{\Big<}}
\def\ra{{\Big>}}

\newcommand{\eq}{\begin{equation}}
\newcommand{\eqx}{\end{equation}}
\newcommand{\eqn}{\begin{eqnarray}}
\newcommand{\eqnx}{\end{eqnarray}}
\newcommand{\f}[2]{\frac{#1}{#2}}

\newcommand{\lm}{\lambda}
\newcommand{\al}{\alpha}
\newcommand{\eps}{\varepsilon}
\newcommand{\bt}{\beta}
\newcommand{\dl}{\delta}

\newcommand{\gm}{\gamma}
\newcommand{\EE}{{\cal E}}
\newcommand{\sn}{\mbox{\rm sn}}
\newcommand{\qqqq}{\quad\quad\quad\quad}
\newcommand{\res}{\mbox{\rm Res}}
\newcommand{\im}{\mbox{\rm Im }}
\newcommand{\arctanh}{\mbox{arctanh}}
\newcommand{\sech}{\mbox{sech}}
\newcommand{\lra}{\longrightarrow}

\begin{document}

\twocolumn[\hsize\textwidth\columnwidth\hsize\csname @twocolumnfalse\endcsname

\title {Prompt Multi-Gluon Production in High Energy Collisions\\
 from Singular Yang-Mills Solutions}
\author {Romuald A. Janik$^{b,c}$, Edward Shuryak$^a$ and Ismail Zahed$^a$}
\address {
$^a$ Department of Physics and Astronomy, State University of New York,
     Stony Brook, NY 11794\\
$^b$ The Niels Bohr Institute,
Blegdamsvej 17, DK-2100 Copenhagen, 
Denmark\\
$^c$ M. Smoluchowski Institute of Physics, 
Jagellonian University,
Reymonta 4, 30-059 Cracow, Poland
}

\date{\today}
\maketitle
\begin{abstract}
We study non-perturbative parton-parton scattering in the
Landau method using singular O(3) symmetric solutions to
the Euclidean Yang-Mills equations. These solutions combine
instanton dynamics (tunneling)  and overlap (transition) between 
incoming and vacuum fields. We derive a high-energy solution
at small Euclidean times, and assess its susequent escape and decay into gluons
in Minkowski space-time. We describe the spectrum of the
{\it outgoing} gluons and show that it is related through a
particular rescaling to the Yang-Mills sphaleron explosion studied
earlier. We assess the number of {\it incoming} gluons in
the same configuration, and argue that the observed scaling
is in fact more general and describes the energy dependence
of the spectra and multiplicities at {\it all} energies.
Applications to hadron-hadron and nucleus-nucleus collisions 
are discussed elsewhere.
\end{abstract}
\vspace{0.1in}
]
\newpage

\section{Introduction}\label{intro}

Tunneling phenomena  related with topology of Yang-Mills fields are
described semiclassically by instantons \cite{BPST,tHooft}. Some manifestations
of these effects related with explicit breaking of U(1) and
spontaneous breaking of  $SU(N_f)$ chiral symmetries are  by now
understood in significant detail due to
strong ties to hadronic phenomenology and  lattice studies, see
e.g.\cite{SS_98} for a review. 

We  know much less about the role of instantons in cross sections of 
various high-energy reactions. Such studies got considerable attention
in the early 90's, when { baryon-number violating} processes through
instantons in the standard model have been actively
discussed~\cite{electroweak,ZMS}. These studies were extended to hard QCD processes 
with small-size instantons, and there are still attempts to see their 
effects on multi-jet production at HERA (for recent review  see \cite{Ringwald}).

Recently~\cite{sz01,KKL} it was suggested that 
nonperturbative configurations composed of instanton/antiinstanton
play an important role in parton-parton scattering amplitudes at
high-energy, and may account for most of the soft pomeron slope and
intercept. The logarithmic rise of the inelastic cross-section
was shown to follow from coherent multiple gluon production as described  
by the semiclassical field following from an interacting instanton-antiinstanton
configuration. This mechanism was shown to be the same for $pp$ and
$\overline{p}p$ \cite{sz01}: so no odderon appears in the classical
limit.  At low energy transfer in the center of mass, the interaction
is dipole, and accounts for the rise in the partial cross section from
first principles. At intermediate and large energy transfer, the dipole
approximation is not reliable as strong instanton-antiinstanton
interactions set in together with unitarity constraints~\cite{ZMS,sz01}. 

It was realized then that unlike perturbative processes, for which
production of each subsequent gluon is associated with a power of
small coupling constant, in instanton-induced processes the cross
section rises with a number of produced gluons reaching a maximum at
some parametrically large value $N\approx 1/\alpha(\rho)$. The physical
reason for this  is that specific {coherent clusters} of gauge
field are produced instead of independent gluons.
In a recent paper~\cite{OCS} the properties of {\em minimal} clusters of such
kind were discussed: the clusters themselves were identified via
constrained minimization of the Yang-Mills energy, with fixed size and 
Chern-Simon number, and their subsequent real time evolution have been
studied both numerically and analytically.

However,  in real collisions the Yang-Mills energy or action
of the final state is only one of the contributing factors.  Another
crucial factor is the {overlap}  between the initial system of colliding
gluons and the instanton, or whatever the tunneling path is. 
A particularly useful way~\cite{khlebnikov} to treat {\it both}
semiclassical factors from first
principles together, and  enforce the unitarity constraints  is to use a
semiclassical approximation
to the partial cross section based on an adaptation of the Landau 
formula for overlaping matrix elements in terms of
singular field configurations.  The occurence of a
singularity is an essential feature of the Euclidean field configuration
that interpolates between the vacuum at $t=-\infty$ with zero energy,
and the escape point at $t=0$ with finite energy.

Diakonov and Petrov~\cite{DP} used this approach 
(Landau method) to rederive the low-energy limit of the amplitude, and
obtain a new result in the high-energy limit. Although they have not
worked out the gauge configuration at small times explicitly, they were
still able to assess the corresponding action, tunneling time and
energy, to predict the behavior of the pertinent cross section at
high-energy. A comparison of the low- and high-energy limits, has 
confirmed that the cross section has a maximum near the sphaleron
energy with a value that is close to the square-root of the cross
section near zero-energy.

The aim of the present paper is to analyze further the singular gauge 
configurations at the {escape time} $t=0$. Following~\cite{DP} we start
with the high-energy limit and show how one can find the gauge
configuration at the {\em turning point} which is the produced gluon cluster
at $t=0$. It turns out that this configuration can be related to minimal YM
sphalerons~\cite{OCS} by a particular {\em scaling law}, containing a
power of $Q/M_s$. We show that the subsequent evolution (explosion) in
real time of these clusters can also be derived, thereby  generalizing 
recent work~\cite{OCS} by our rescaling.  The resulting spectrum of gluons and
their multiplicity are readily obtained. We argue that the
prescribed scaling law is valid not only at high energies, above the
sphaleron mass, but in fact at $all$ energies.

In section 2 we introduce the standard notations for
the spherically symmetric gauge configurations used.
In section 3, we recall the key points in assessing 
the inelastic parton-parton scattering cross-section
in the eikonal approximation.  In section 4, we show that the singular gauge
configurations at the escape points follow from the sphaleron point
for all energies through a pertinent { rescaling}, which is our
main result. In sections 5, we assess
the number of incoming and outgoing gluons per sphaleron for fixed center of mass
energy in the semiclassical approximation. Our conclusions are in
section 6.
The Appendices contain a number of useful results including
a stability analysis of the escaping configurations under perturbative
light pair-decay.

\section{O(3) Symmetric Yang-Mills}

We consider the QCD Yang-Mills theory wherein all dimensions 
are rescaled away by the sphaleron (antisphaleron) mass $M_S$, 

\be
M_S= \frac 1{4\alpha} \,\int_0^\infty \,dx\,x^2\frac {96\rho^4}{(x^2+\rho^2)^4}
=\frac {3\pi}{4\,\alpha\,\rho}\,\,,
\label{S16}
\ee
unless specified otherwise. In the vacuum $\alpha\approx 0.3$, $\rho\approx
1/3$ fm with typically $M_S\approx 3$ GeV. In the scattering process,
the sphaleron (antisphaleron) size $\rho$ may change. We work mainly in temporal gauge.
The YM potential ${\bf V}$, kinetic energy ${\bf K}$ and 
Chern-Simons number  ${\bf N}$  are

\be
&&{\bf V} = \frac{M_S}{4\pi\,\alpha}\,
\int d\vec{x}\,\frac 14\,(F_{ij}^a)^2\nonumber\\
&&{\bf K} = \frac{M_S}{4\pi\,\alpha}\,
\int d\vec{x}\,\frac 12\,(\dot{A}_i^a)^2\nonumber\\
&&{\bf N} = \frac 1{16\pi^2}\,
\int d\vec{x}\,\epsilon_{ijk}\,(A_i^a\,\partial_j\,A_k^a
+\frac 13 \epsilon_{abc}\,A_i^a\,A_j^b\,A_k^c)\,\,,
\label{S01}
\ee
where we have ignored quarks for simplicity.

In parton-parton scattering with large $\sqrt{s}$,
the incoming kinematics boils down to an eikonalized cross section 
as in (\ref{S1}) with a partial cross section $\sigma\,(Q)$ with
$Q\approx M_S\ll \sqrt{s}$. In the center of mass it is reasonable
to consider the gluonic configurations that maximizes $\sigma\, (Q)$
to possess higher symmetries than completely arbitrary fields. Here we
take them to have spherical $O(3)$ symmetry of the sphaleron

\be
A_i^a (\vec{x}, t) = &&+ \epsilon_{aij} \,n_i\,(1-A(x,t))/x\nonumber\\
&&+(\delta_{ai}-n_an_i)\,B(x,t)/x\nonumber\\
&&+n_an_i\,C(x,t)/x
\label{S8}
\ee
where $n_i=\vec{x}_i/x$ is a unit 3-vector, and $|\vec{x}|=x\geq 0$
a radial variable. Since we are interested in singular gauge fields,
we assume the three 2-dimensional independent functions $A,\,B,\,C$
to be continuous and differentiable everywhere in Euclidean space,
except at $x=0$ where
a singularity will be located for fixed times $\pm T/2$. In terms of 
(\ref{S8}) the Euclidean action 

\be
S=\int dt\, ({\bf K} + {\bf V}) 
\label{S02}
\ee
reads

\be
&&S=\frac 1{\alpha} \int\, dt\,\int_0^\infty\,dx
( \dot{A}^2 +\dot{B}^2+\frac 12 \,\dot{C}^2 + {A'}^2 + {B'}^2 \nonumber\\
&& +\frac{(A^2+B^2-1)^2}{2x^2} + \frac{C^2 (A^2+B^2)}{x^2} + 
\frac {2C(A'B-AB')}x)
\label{S9}
\ee
where the time variable has been rescaled through $t\,M_S\rightarrow t$,
in (\ref{S9}). The integration interval is to be specified below in the
presence of a time-singularity. Note that the energy

\be
-Q= {\bf K} -{\bf V}
\label{S03}
\ee
where in Euclidean space $-{\bf V}$ plays the role of the potential.
For self-dual configurations $Q=0$. The Chern-Simons number is

\be
{\bf N} = \frac 1{2\pi} \int_0^\infty \,dx\,
\left(A'B-B'A+ \frac Cx (A^2+B^2-1)\right)\,\,.
\label{S10}
\ee
The dual to the O(3) ansatz used here that maps on the antisphaleron
follows similar reasoning and results as shown in Appendix A. The
difference is a negative ${\bf N}$.

\section{The Evolving Views on Non-perturbative High Energy  Processes}

In this work we discuss semi-hard
parton-parton processes at fixed $\sqrt{-t}\sim 1 \, {\rm GeV}\ll \sqrt{s}$, 
using nonperturbative QCD gauge configurations related to topological
tunneling. The original approach \cite{electroweak} wich we followed
 in~\cite{sz01} was based on semiclassical instanton solution in the amplitude.
To leading order in the small tunneling or diluteness factor of
the typical density and size of instantons in the vacuum,
$\kappa=n_{\rm inst}\rho^4\approx 10^{-2}$ (see e.g.\cite{SS_98} for details), 
the inelastic cross section reads

\be
\sigma_{\rm in} (s,t) \approx &&\,{\bf C}\,\pi\,\rho^2\,{\rm ln \,s}\,
\int\,dq_{1\perp}\,dq_{2\perp}\,
\,{\bf K} (q_{1\perp} , q_{2\perp} , t) \nonumber\\&&\times\,
\int^\infty_{(q_{1\perp}+q_{2\perp})^2}\, dQ^2\,\kappa^2\,{\bf B}(Q)
\label{S1}
\ee
where ${\bf K}$ is a pertinent
instanton form factor at fixed $-t\ll s$, containing through-going
partons in the form of Wilson lines.  $\kappa$ appears squared 
since the cross section involves the squared amplitude.
{\bf B}(Q) is the partial multi-gluon cross section for fixed $Q^2\ll s$ and 
${\bf C}$ an overall constant which accounts for both the instanton
and antiinstanton contributions to the initial state. To exponential 
accuracy

\be
\kappa^2{\bf B} (Q)\approx  {\rm Im} \int\, dT\, e^{QT-{\bf S}(T)}
\label{S2}
\ee
where ${\bf S} (T)$ is the effective action describing instanton-antiinstanton 
interaction for a time-separation $T$, which is defined to reduce to 
 twice the free instanton action at large $T$. This effective action, also known as 
the ``holy grail function'', is supposed to sum up contributions of {\it
any} number of produced gluons. For small $Q$ or large $T$,
the dipole approximation is valid and (\ref{S2}) rises exponentially
with $Q$~\cite{electroweak}. However, as emphasized first by Zakharov \cite{ZMS}
the  unitarity constraints on the partial cross section
requires ${\bf B} (Q)$  to fall at large $Q$. Shifman and Maggiore \cite{ZMS}
 argued that the unitarization could be qualitatively
enforced by resumming chains of instantons and antiinstantons.
In \cite{sz01} we followed this idea and indeed
found that the dominant 
contribution to (\ref{S2}) occurs at the sphaleron point
for which ${\bf B} (M_S) \approx 1/\kappa$. Hence,

\be
\sigma_{\rm in} (s,t)\approx &&{\bf C}\,\pi\rho^2\,\kappa\,{\rm ln \,s}\,\,
\int\,dq_{1\perp}\,dq_{2\perp}\,
\,{\bf K} (q_{1\perp} , q_{2\perp}, t)\,\,.
\label{S3}
\ee
The rise in the partial inelastic cross section due to
the production of the Yang-Mills (YM) sphaleron~\footnote{Of course, there is
also production of the YM antisphaleron, which carries
opposite Chern-Simons number.}, results into an increase of the
inelastic cross section by one power of the diluteness factor
$\kappa$,  or about a 100-fold increase~\cite{sz01}. 

This qualitative solution of the problem implies however that the
maximal cross section does not correspond to a well-separated
instanton-antiinstanton pair, but rather to a close pair with
$T\approx \rho$ whereby half of the asymptotic action is annihilated.
Detailed studies of close instanton-antiinstanton configurations
of such kind have been made recently in~\cite{OCS}. An important new
element of that paper was the emphasis on the t=0 3-plane, treated as a
unitarity cut and explicitly describing the turning (escaping to
Minkowski space-time) gauge field configuration.  Although this paper
was not dealing with the cross section calculation, following its logic
one should e.g. modify the form-factor {\bf K}  in (\ref{S1}) by using
Wilson lines in a half-instanton-antiinstanton field as described for
instance by the Yung ansatz.

\begin{figure}[h]
\epsfxsize=6cm 
\epsfbox{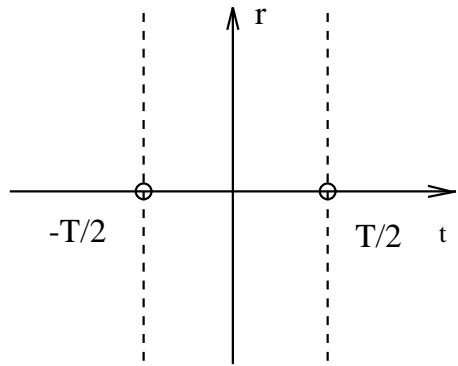}
\caption{\label{fig_plane}
Euclidean pace-time (r,t) plane. The circles indicate the location
of the gauge singularities, and the vertical lines show where the
0-energy and Q-energy solution are joined.}
\end{figure}

However, even this solution of the problem would not be ideal, as it
treats independently two small semiclassical factors, the
instanton-antiinstanton interaction {\bf S} in the vacuum and
the form-factor {\bf K}. The idea behind the Landau method 
is to treat these semiclassical contributions simultaneously. The
essentials of the method are as follows (more details can be found in
\cite{DP}): In quantum mechanics the overlap between the ground 
state and a highly excited state can be rewritten as a difference
of two shortened actions for a pair of paths with energy 0 and $Q$
respectively. The corresponding integrals interpolate between
the turning points and the singularity at infinity.
The location of the singularity of the gauge configuration plays the role of such
an infinity. Specific Euclidean paths, used originally in~\cite{khlebnikov,DP}, have
singularities in the $A_\mu$ field  located at r=0 and time $\mp T/2$, 
see fig.\ref{fig_plane}. Outside the region marked by the dashed lines the 
solution is the universal singular instanton, describing the ground
state. Between the dashed lines it is supposed to be the (so far missing)
energy-Q solution: both solutions join smoothly at the dashed lines. 
Our aim is to find the interpolating solution, at least approximately, 
and investigate at $t=0$ the internal field
structure of relevant turning states. Their subsequent evolution will
eventually lead to predictions of what is actually produced in the collision.

\section{More details on Singular Yang-Mills}

\subsection{Singular fields: $|t|>T/2$}

The branch of the
field that interpolates between the vacuum at $t=-\infty$ and the
singularity at $t=-T/2$ with zero energy and minimizes (\ref{S02})
is an instanton. The conjugate branch is an an antiinstanton and
interpolates between the singularity at $t=+T/2$ and the vacuum at
$t=+\infty$. In covariant gauge, both branches are given by

\be
A_\mu^a (x (\pm) ) = &&2\,\eta_{\mu\nu} (\pm)\,x_\nu (\pm) 
\Phi (x^2(\pm))\nonumber\\ = && 
\frac{2\,\eta_{\mu\nu}^a (\pm ) \,x_\nu (\pm )\,\rho^2}
{(\rho^2-x^2(\pm))\,x^2(\pm)}
\label{S4}
\ee
where the $-$ field refers to $t<-T/2$ time-branch, and
the $+$ field refers to the $t>T/2$ time-branch, with

\be
x(\pm)_\nu = (\vec{x},t_\pm)_\nu =
(\vec{x}, t\mp T/2 \pm \rho)_\nu\,\,.
\label{S5}
\ee
The t'Hooft symbols are identified as $\eta(+)=\eta$
(antiinstanton) and $\eta (-)=\overline\eta$ (instanton).
For $|t|>T/2$ (\ref{S4}) are  singular self-dual solutions
to the QCD Yang-Mills equations with zero-energy. Note that
the singularity in (\ref{S4}) stems from the change $\rho\rightarrow i\rho$
in the self-dual O(4) instanton. $A(+)$ is the time-conjugate of $A(-)$.

The axial gauge $A_4=0$ is commensurate with the
$O(3)$ symmetry, and the results for the $\pm$ branches
follow by using the hedgehog gauge transformation 

\be
U(\vec{x},t) = {\rm exp}\left(
i\vec{x}\cdot\vec{\tau}\,\int^t\,dt'\,\Phi ({\vec{x}}^2+{t'}^2)\right)\,\,,
\label{S6}
\ee
modulo static gauge transformations. Under the action of (\ref{S6}) 
the $A_4$ part in the Lorentz gauge (\ref{S4}) is gauged to zero.
The residual static gauge transformations are fixed by 
fixing the positions of the singularities in the axial-gauge
to coincide with those in the Lorentz gauge at $t=\mp T/2$. In
particular,

\be
A_i^a (\vec{x},-T/2) =&& A_i^a (\vec{x},+T/2)\nonumber\\
 =&& - (\epsilon_{aij} \,x_j +
\delta_{ai}\rho)\,\, \frac{2\rho^2}{x^2 (x^2+\rho^2)}\,\,.
\label{S7}
\ee

\subsection{Singular fields: $|t|<T/2$}

The gauge configuration in the time interval $|t|<T/2$,
follows from the YM equations using (\ref{S7}) as the 
singular boundary conditions. They are no longer constrained
by self-duality, and hence carry finite energy $Q$. Fixed $Q$
relates to fixed $T$ through $Q=dS/dT$ where $S$ is the Yang-Mills 
action in the time interval $|t|\leq T/2$.

\subsubsection{Above the Sphaleron}

For small Euclidean times $T/\rho\ll 1$ or large energy, the singular 
boundary conditions (\ref{S6}) common to both Lorentz and axial gauge,
imply for the axial gauge decomposition (\ref{S8}) that in this limit
$B\rightarrow C \gg A$ throughout~\cite{DP}. In this limit, we will
call  $D=2B=2C$
the action (\ref{S9}) reduces to

\be
S\approx \frac 3{4\alpha}\int_{|t|<T/2}dt \int_0^\infty\, dx\,
\left( \frac 12\,\dot{D}^2 +\frac 1{8x^2}\,D^4\right)\,\,,
\label{XS1}
\ee
the energy is

\be
-\frac Q{M_S}\approx +\frac 3{4\alpha} \int_0^\infty\, dx\,
\left( \frac 12\,\dot{D}^2 - \frac 1{8x^2}\,D^4\right)\,\,,
\label{XS2}
\ee
the Chern-Simons number is

\be
{\bf N} \approx  \frac 1{2\pi} \int_0^\infty \,dx\, \frac {D^3}{8x}\,\,,
\label{XS3}
\ee
with the boundary condition $D(r, \pm T/2) = -{4\rho}/r$.
The action (\ref{XS1}) is extremal for 

\be
\frac{D(r,-T/2)}{2r/(t+T/2)} =
\int_1^{D(r,t)/D(r,-T/2)}\,\frac{dx}{\sqrt{x^4-1}}\,\,.
\label{XS4}
\ee
The transcendental equation (\ref{XS4}) can be solved
numerically. The solution is shown in Fig.~1 (thick line)
for $\rho/T=10$.

A good approximation at the escape point is 

\be
D(r, 0) \approx  \frac {4\rho \,r}{r^2+ \frac{\sqrt{2}}{K}\,\rho\,T}
\label{XS5}
\ee
which interpolates exactly between the asymptotics of
the transcendental solution with $K=1.854$. (\ref{XS5}) 
is also shown in Fig.~1 (thin line). Its corresponding
initial radial density is

\be
\Theta_{00} (r,0) \approx  \frac {4\pi}{g^2} \,
\frac{24\,\rho^4\,r^2}{(r^2+ (M_S/Q)^{2/5} \rho^2)^4}\,\,,
\label{XS6}
\ee
which integrates to $Q$. Note that 
the tunneling duration $T$ relates to 
the energy $Q$ through

\be
\frac T\rho = \frac K{\sqrt{2}}\,\left(\frac {M_S}{Q}\right)^{2/5}\,\,.
\ee
The Chern-Simons number at the turning point for (\ref{XS5}) is

\be
{\bf N} =\frac 12 \, \left(\frac {Q}{M_S}\right)^{2/5}\,\,.
\label{XS7}
\ee
At the sphaleron point with $Q=M_S$ the configuration (\ref{XS5})
carries Chern-Simons number ${\bf N}=1/2$: it is a sphaleron.

\begin{figure}[h]
\epsfxsize=8cm 
\epsfbox{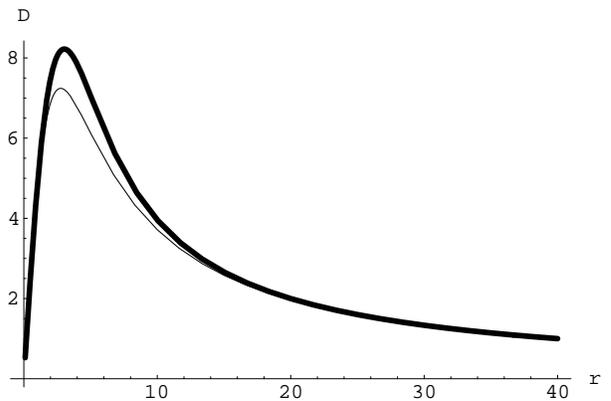}
\caption{$D(r,0)$ for $\rho/T=10$. See text.}
\end{figure}
 
For $Q>M_S$ the initial configuration follows from the sphaleron by
a simple rescaling of the size $\rho$ and the energy density,

\be
&&\rho\rightarrow \rho/\lambda\nonumber\\
&&\Theta_{00}\rightarrow \lambda^4\,\Theta_{00}
\label{XS8}
\ee
with $\lambda=(Q/M_S)^{1/5}$. 
In section {\bf VB} we will evolve the gauge field configuration into
Minkowski space using Luscher-Schechter (LS) solutions~\cite{LS}. These solutions
have a purely magnetic field configuration at $t=0$ and the (radial)
energy profile 
\eq
\Theta_{00}(r,0)=4 \pi r^2 \Theta_{00}(\vec x,0)= \f{4\pi}{g^2} 
\f{48 \epsilon \rho_{LS}^4 r^2}{(r^2+\rho_{LS}^2)^4}\,\,.
\eqx
Comparing with the rescalings (\ref{XS8}) we see that we have to use
the LS solutions with the parameters
\eqn
\rho_{LS}&=&\f{\rho}{\lm} \nonumber\\
\epsilon &=&\f{\lm^4}{2}\,\,.
\label{LSPAR}
\eqnx



\subsubsection{Below the Sphaleron}

Below the sphaleron the analytical analysis is more involved
in general. For small energy $Q$ or large times $T$, a perturbative
expansion around the singular instanton-antiinstanton configuration
has been carried out in \cite{DP}. As we will argue below,
the multiplicities below the sphaleron follow from the same 
rescaling (\ref{XS8}) with $Q<M_S$.

\section{Gluons In/Out}

In this section we estimate the number of incoming (virtual) and outgoing
(real) gluons present in the semiclassical singular gauge configurations
for arbitrary parton-parton center of mass energy $Q$.

\subsection{Incoming Gluons}

The number of incoming gluons follow from the exact Euclidean solutions
at large Euclidean times by expanding in powers of the space Fourier 
transform of the large Euclidean asymptotics of the singular fields
(\ref{S4}) as in \cite{khlebnikov,khl2}. In Lorentz gauge, 

\be
{\cal A}_0^a (t,\vec{k}) = &&2i\,\hat{k}_a\, {\cal Q} (\pm,k)\,e^{-kt}\nonumber\\
{\cal A}_i^a (t,\vec{k}) = &&-2\,\hat{k}_a\,\hat{k}_i\,{\cal Q} (\pm,k)\,e^{-kt}
\nonumber\\
&&+\frac{\lambda_i^m}{\sqrt{2k}}\,f_a^m (\pm,k)\,e^{-kt}
\label{IN1}
\ee
with,

\be
{\cal Q} (\pm,k) = \frac{2\pi^2}g\,\rho^2\,e^{\pm k\,(T/2-\rho)}
\label{IN2}
\ee
and 

\be
&&f_a^m (\pm,k) = \nonumber\\
&&\frac {4\pi^2\,\rho^2}{g}\,\sqrt{2k}\,
\left(-\lambda_a^m +i\epsilon_{abj}\,\lambda_b^m\,\hat{k}_j\right)\,
e^{\pm \,k(T/2-\rho)}\,\,.
\label{IN3}
\ee
The transverse polarizations are denoted by $\lambda_a^m$. In terms
of the Fourier components (\ref{IN3}) the density of incoming 
{ transverse} gluons is proportional to the occupation number
$\overline{a}\,a$ with

\be
a_i^m (\theta, k) = f_i^m (+, k)\,e^{-k (T-\theta)/2}\,\,,
\label{IN4}
\ee
where $\theta$ is a parameter fixed by requiring the total energy of
the incoming gluons to match the energy $Q$~\cite{khlebnikov,khl2}. 
We note that $T$ drops in the combination (\ref{IN4}). Hence, the
density of transverse gluons per unit wavenumber is

\be
{\bf n}(k) =
\frac{16\pi}{\alpha}\,\rho\,(k\rho)^3\,e^{-4k\rho\,(M_S/2Q)^{1/5}}\,\,,
\label{IN5}
\ee
and the corresponding energy density per wavenumber is

\be
{\omega} (k) = k\, {n}(k) = \frac{16\pi}{\alpha}\,(k\rho)^4\,
e^{-4k\rho\,(M_S/2Q)^{1/5}}\,\,.
\label{IN5x}
\ee
Under the rescaling (\ref{XS8}) the energy density (\ref{IN5x})
of the incoming transverse gluons follows from the
sphaleron point and integrates to $Q$. The virtual number of
 gluons stripped by a sphaleron is

\be
N_{{\rm in}} (Q) = \frac{3\pi}{8\alpha}\,\left(\frac{2Q}{M_S}\right)^{4/5}\,\,,
\label{IN6}
\ee
and similarly for the antisphaleron. Each virtual gluon carries away
about $M_S/3\approx 2/3$ GeV. We recall that these gluons 
are absorbed from the 2 eikonalized partons involved in the inelastic 
cross section (\ref{S1}).

\subsection{Outgoing Gluons}

To assess the number of outgoing transverse gluons produced by the
singular gauge configurations in the semiclassical approximation,
we need to know the gauge configurations at the escape point and
their further Minkowski time-evolution, much like the decay of the
sphaleron in the standard model~\cite{EW}. 
The escaping sphaleron in Minkowski space is related to
an analytical solution discovered by Luscher and Schechter~\cite{LS} 
(LS) as discussed in~\cite{OCS,PAST}. What is remarkable 
in our case, is that through the scaling laws (\ref{XS8}) we have tied
features of this solution (energy density and multiplicity) to those
of the escaping singular Yang-Mills fields above the sphaleron point.

\subsubsection{At the Sphaleron}

The LS solution in Minkowski space at the sphaleron point
can be expressed in terms of
elementary functions~\cite{ACTOR} (some
helpful relations can be found in the Appendices).
It is strongly peaked 
around the light cone $t\sim r$ as it travels luminally,
and for large times $t\sim r\gg \rho$ in covariant gauge
it simplifies 

\be
A_0^a (t,v) \sim  && -\frac {F'}g\,\,n_a\nonumber\\
A_{i}^a (t,v) \sim  && +\frac {F'}{g} \, n_a\,n_i \nonumber\\ &&
+ \frac{F'}{gt} \,(v\,(\delta_{ai}-n_an_i) + \rho
\,\epsilon_{iaj}\,n_j)\,\,,
\label{LS1}
\ee
with $v=r-t$ and

\be
F'(v) =&&\frac{2\rho\,f}{\rho^2+v^2}\nonumber\\
f(v) =&&\frac 12 \left(1-\frac{\sqrt{2}}{{\rm ch}\,(\sqrt{2}\,\xi)} \right)\,\,,
\label{LS2}
\ee
and $\xi = {\rm tan}^{-1}(\rho/v)$. Note that
$f(\pm\infty)=1/2-1/\sqrt{2}\sim -0.207$, with 
$f(0)>f(\infty)$. Choosing $F(0)=0$ we see that
$F(v)$ is an odd function with asymptotics
$F(\pm\infty)=\pm 0.216$.  
We discuss some more details of the solution in Appendix A, and show
that behavior of gauge invariants is the same as found in \cite{OCS}.

The gluon number and spectrum is a subtle issue,
and should be performed in ``physical'' gauges.
In the covariant gauge,  the large time asymptotics of the temporal 
and longitudinal gauge fields fulfill $\partial\cdot A=0$. They are
proportional to $F'(v)$ with support only on the light-cone. 
This is a gauge artifact in the covariant gauge, and
can be removed by transferring to temporal gauge 
by using the hedgehog gauge transformation

\be
\omega (v) = e^{i\tau\cdot n\,F(v)}
\label{LS4}
\ee
which yields $A_0^a=0$.
The temporal gauge is canonical in the
sense that Gauss law is easily implemented by restricting
to the transverse gluons, and the vacuum state is normalizable.
In this gauge the large time asymptotics of the field is
purely transverse and falls as $1/t\sim 1/r$,

\be
&&{A}_i^a (t,v) \sim 
\frac 2{gt}\,({\rm sin^2}F +\frac {F'}2 (\rho\,{\rm cos}\,(2F) + v
{\rm sin}\,(2F) ))\,\epsilon_{aij}\,n_j\nonumber\\
&&+\frac 1{gt}\,(-{\rm sin}\,(2F) + F'\,(\rho\,{\rm sin}\,(2F)-v
{\rm cos}\, (2F) )\,(\delta_{ai}-n_a\,n_i)\,\,.\nonumber\\
\label{LS5}
\ee
Note that the gauge transformation (\ref{LS4}) modifies the Chern-Simons
number (\ref{S10}). 
The transverse fields in both covariant and axial gauge weaken asymptotically
as $1/t$. Asymptotically the Yang-Mills solution originating from the
sphaleron point Abelianizes, thereby allowing for a free wave
interpretation. It is easier to carry the analysis in covariant 
gauge~\footnote{In temporal
gauge there is a subtlety related to the constant modes which do not admit
a spectral representation. Indeed, we have checked that the space
Fourier component of (\ref{LS5}) exhibit a non-spectral term
$t\,\delta (k)$.}.

The large time transverse asymptotic of (\ref{LS1}) fulfills trivially
the covariant gauge condition, and admits a normal mode decomposition 
in the form

\be
&&{\cal A}_i^a (t, \vec{k}) =\frac{(2\pi)^{\f{3}{2}}}{\sqrt{2k}}\nonumber\\
&&\times 
\left(\lambda_i^m (\vec{k})\,b^{am} (\vec{k}) e^{-ikt} + 
\lambda_i^m (-\vec{k})\,b^{am\,*} (-\vec{k}) e^{+ikt}\right)
\label{LS6}
\ee
with the $\lambda$'s as the 2 real polarizations,

\be
&&\lambda_i^m (\vec{k})\,b^{am} (\vec{k}) = \frac{\rho}{g\sqrt{\pi\,k}}\nonumber\\
&&\times (-i\epsilon_{aij}\hat{k}_j \,J(k\rho) + (\delta_{ai}-\hat{k}_a\hat{k}_i)
\,J'(k\rho))\,\,,
\label{LS7}
\ee
and

\be
J(k\rho) = 2\,{\rm Re}\,\int_0^{\pi/2}\,dy\,e^{ik\rho\,{\rm cotan}\,y}\,
\left(1-\frac{\sqrt{2}}{{\rm cosh}(\sqrt{2}\,y)}\right)\,\,.
\label{LS8}
\ee
The transcendental function (\ref{LS8}) is not emmenable to
closed form, but is well behaved for $k\rho\ll 1$

\be
J(k\rho) &\sim& \pi-4 \arctan \left(\tanh\left(\f{\pi}{2\sqrt{2}}
\right) \right) \\
J'(k\rho) &\sim& \pi\,(\sqrt{2}-1)
\ee
%

In terms of the normal mode decomposition (\ref{LS6}), the
asymptotic density of transverse gluons is proportional to the 
occupation number $|\lambda\cdot b|^2$ of the transverse modes,

\be
{\bf n} (k) =4\pi\,k^2\,|\lambda\cdot b|^2= \frac{8\rho^2k}{g^2}\,
\left( J^2 (k\rho) + {J'}^2(k\rho) \right)\,\,.
\label{LS9}
\ee
The number of gluons with small energy grows as $k\rho$, while the
number of gluons with high energy falls as $k\rho\,e^{-2\,k\rho}$.
The total number of prompt gluons emitted by a sphaleron is 

\be
N_{\rm out} (M_S)
=\,\frac 8{g^2} \,\int_0^\infty\, x\,dx\,(J^2(x) + {J'}^2 (x) )=
\frac{1.1}{\alpha} 
\label{LS9xx}
\ee
where the~last integration has been performed numerically\footnote{
The present spectrum is very similar to the one obtained numerically
in~\cite{OCS} using the temporal gauge.  It is different from the one discussed 
analytically in~\cite{OCS} using a {different} gauge where the gluon number
was found to diverge logarithmically at small $k$.
The gluon number is also found to diverge in axial gauge in relation to the
$k=0$ modes (see footnote 1). The results discussed above are free from
gauge artifacts.}. The same number of gluons are produced
through the antisphaleron. Each mode
in the transverse asymptotic (\ref{LS6}) is normal, so that the energy
density carried by these modes is $\omega (k) =k\,{\bf n} (k)$, which
integrates to give back the sphaleron mass by energy conservation,

\be
M_S = \frac{8}{g^2\rho}\int_0^\infty\,dx\,x^2 (J^2(x)+{J'}^2(x))
=\f{2.2}{\alpha\rho} \,\,.
\label{LS10}
\ee
The numerical result (\ref{LS10}) is off by $7\%$ in comparison to
the exact sphaleron mass (\ref{S16}) which is a measure of the onset
of the asymptotic normal mode expansion (\ref{LS6}).
In Appendix C we show that the expanding sphaleron configuration 
is stable under pair production of light quarks and gluons.

\subsubsection{Away from the Sphaleron by Rescaling}

The escape configurations above the sphaleron follows from the gauge
configuration (\ref{XS4}-\ref{XS5}). The latter yields the energy
density of the sphaleron after the rescaling (\ref{XS8}) and 
a Chern-Simons number of 1/2 at the sphaleron point. Since the 
electric field vanishes at the escape point, we conclude that 
it is very plausible that this gauge configuration is gauge equivalent
to the LS gauge configuration with the parameters (\ref{LSPAR}).

The analysis of the preceding section may be performed just with the
substitution of $f(v)$ by
\eq
\f{1}{2}\left( 1-\sqrt{1+\sqrt{2\epsilon}} \cdot \mbox{\rm dn}\left(
\sqrt{1+\sqrt{2\epsilon}}\, \xi,\f{1}{m} \right)\right)
\eqx
for $\epsilon<1/2$, and with
\eq
\f{1}{2}\left( 1-\sqrt{1+\sqrt{2\epsilon}} \cdot \mbox{\rm cn}\left(
(8\epsilon)^{\f{1}{4}} \xi,m \right)\right)
\eqx
for $\epsilon>1/2$, where
\eq
m=\f{1+\sqrt{2\epsilon}}{2\sqrt{2\epsilon}}
\eqx

In Fig.~2 we show the multiplicity distributions
for various values of $\lambda$.
Numerically there is not much difference between the solutions
obtained from the elliptic LS profiles and an appropriate rescaling
(\ref{XS8}) of the sphaleron results.
In particular,
\be
{\bf n} (k) \sim \lambda^4\,\frac{8k\rho^2}{g^2\lambda^2}\,
\left( J^2 \left(\frac{k\rho}{\lambda}\right) + 
{J'}^2\left(\frac{k\rho}{\lambda}\right) \right)\,\,,
\label{LS10x}
\ee
with $\lambda=(Q/M_S)^{1/5}$.
The total number of prompt gluons emitted
above the sphaleron is (in this approximation)

\be
\frac{N_{\rm out}(Q)}{N_{\rm out} (M_S)} = \left( \frac{Q}{M_S}\right)^{4/5}\,\,.
\label{LS11}
\ee
The ratio of in (virtual) to out (real)  gluons per sphaleron is a number 
independent of $Q$: $N_{\rm in}/N_{\rm out} \sim 2$. Prompt inelastic scattering in
QCD is from few-to-few (small $Q$)  or many-to-many gluons (large $Q$). 

\begin{figure}[h]
\epsfxsize=8cm 
\epsfbox{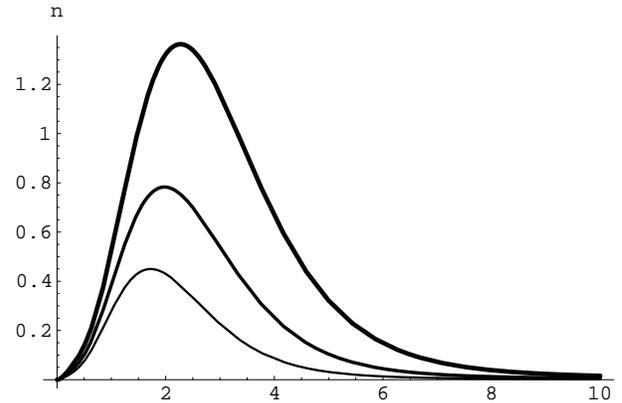}
\caption{Density of emitted gluons per sphaleron (multiplied by
$\alpha/ \rho$) for $\lambda=0.5,1,2$ versus $k\rho$.}
\end{figure}

Below the sphaleron barrier, there are two turning points: $t=-T_1/2$
with zero Chern-Simons number ${\bf N=0}$, and $t=0$ with unit Chern-Simons
number ${\bf N=1}$. The two gauge configurations and thereby
multiplicities are related by the gauge transformation (\ref{S6}). 
The remarkable
similarity between the scaling law (\ref{LS11}) for the outgoing gluons
above the sphaleron point and the scaling law (\ref{IN6}) for the
incoming gluons both above and below the sphaleron point leads us to
{\em conjecture that (\ref{LS11}) holds for the outgoing gluons  below the
sphaleron point as well}.

The total number of prompt gluons emitted by the escaping singular
Yang-Mills configurations below the sphaleron follows the scaling
law (\ref{LS11}), which is seen to vanish at the instanton point.
This result follows from a saturation of the partial cross section via
classical and singular solutions to the Yang-Mills equations, and is different from
the one derived recently using a minimization of the energy at the
escape point by constraining the size and Chern-Simons number at the
escape point~\cite{OCS}. The latter is likely to provide a lower
bound on the partial cross section, while the former saturates it.

\subsection{Averaging Gluons}

In so far, we have considered the production of prompt gluons
 for fixed $Q$ in the inelastic production cross section given
by (\ref{S1}). For the singular Yang-Mills solutions considered
here, the partial cross section $\sigma (Q)$ has been derived
for small and large $Q$ in~\cite{DP}. In units of the 
scaled energy $x=Q/M_s$,
their result, with exponential accuracy (and not too close to the x=1 point)
is

\be
\sigma(x) = \sigma_+(x) \,\theta(x-1) + \sigma_-(x) \,\theta(1-x)
\label{D1}
\ee
with

\be
\sigma_\pm (x) =e^{\frac{4\pi}{\alpha}\,{\bf F}_\pm (x)}\,\,.
\label{D2}
\ee
The (known part of the) `holy-grail' function reads
\footnote{The scaling laws derived above are specific to the energy
density and the gluon multiplicities. They do not carry to the action
density needed to assess semiclassically the partial cross section
(\ref{D1}). For that we need explicitly the escaping classical fields
starting from (\ref{XS4})-(\ref{XS5}) for instance.}

\be
{\bf F}_+ (x) = &&-0.482\,x^{3/5}\nonumber\\
{\bf F}_- (x) = &&-1 + 0.6185\,x^{4/5} \nonumber\\
&&-0.0710\,x^{6/5} +0.0122\,x^{8/5}\,\left(-{\rm ln} \,x+ {\rm
const}\right)\,\,.
\label{D3}
\ee
The first contributions in ${\bf F}_\pm$ are from the semiclassical
singular gauge configurations alone, while the last two contributions
in ${\bf F}_-$ are from the one-loop and two-loop contributions 
respectively~\cite{DP}. The initial increase in the partial cross section 
in (\ref{D1}) follows from the rapid increase in the tunneling rate at
the expense of the decrease in the matrix-element overlap. At the
sphaleron point $x=1$, ${\bf F_+} (1) \sim {\bf F}_- (0)/2$ which is about
half the instanton suppression factor,

\be
\kappa =\sigma_+(1) \sim \sqrt{\sigma_-(0)}\sim\sqrt{e^{-4\pi/\alpha}}\,\,,
\ee
in agreement with the unitarization arguments in~\cite{ZMS,sz01}. The final and
rapid decrease in the cross section passed the sphaleron point is caused
by the decrease in the overlap between the initial and final states of
the inelastic collision process.

In this subsection we would like to show that in the {\em inclusive gluon production}
a sharp peak remains, even after all gluon multiplicity factors are
included. In Fig.~4  we plot $x^{4/5}\sigma(x)$, with the typical
instanton  action in the QCD vacuum set to be
$2\pi/\alpha_s(\rho)=12$. From this figure one can see  that the
r.h.s. of the peak is larger than the l.h.s.: it looks like
Breit-Wigner peaks distorted by multibody phase space.

Using (\ref{D1}) we can assess the averaged number of prompt gluons
produced in a parton-parton scattering process at large $\sqrt{s}$,

as a function of the invariant mass transfer in the t-channel,
with the measure $\mu (x,t)$ fixed by the partial cross section

\be
\mu(x,t) = \sigma(x)/
\int_{-t/M_S^2}^\infty\,dx^2\,\sigma(x)\,\,.
\label{D5}
\ee
$N_{\rm in,out}$ are the number of gluons at the sphaleron mass.
The results for the ratio (\ref{D4}) are displayed in Fig.~5
versus $-t/M_S^2$. The undetermined constant in (\ref{D3}) was
adjusted to $2.418$ to insure a smooth transition at $x=1$.
For $\sqrt{-t}$ within $M_S$, the averaged multiplicity is increased
by only a factor
about 1.1 compared to the multiplicity at the sphaleron mass.

\be
\frac{{N}_{\rm in,out} (t)}{N_{\rm in,out}} =
\int_{-t/M_S^2}^\infty\,dx^2\,\mu(x,t)\,x^{4/5}\,
\label{D4}
\ee

\begin{figure}[hb]
\epsfxsize=6cm
\epsfbox{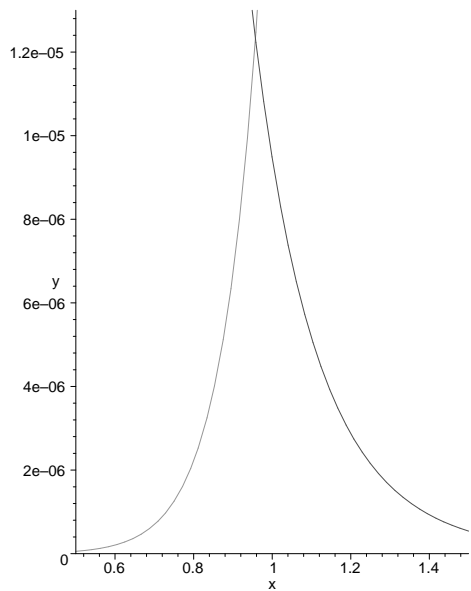}
\caption{The inclusive gluon multiplicity  $x^{4/5}\sigma(x)$ versus 
energy in units of the sphaleron mass $x=Q/M_s$.}
\end{figure}

\begin{figure}[ht]
\epsfxsize=8cm 
\epsfbox{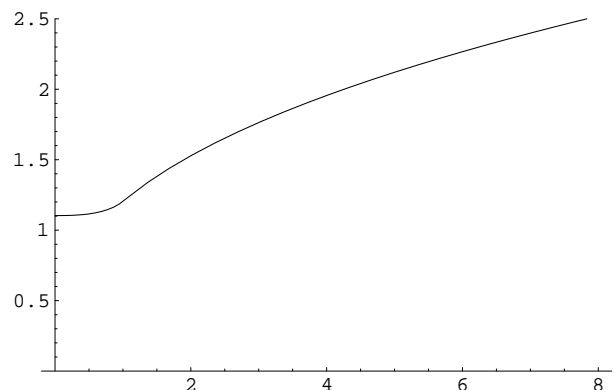}
\caption{The ratio (\ref{D4}) of prompt gluons emitted as a function
of $-t/M_S^2$.} 
\end{figure}

\section{Conclusions}

We have presented a semiclassical assessment of the inelastic cross
section for multi-gluon production through instanton induced 
parton-parton scattering, using singular classical solutions to the
Euclidean Yang-Mills equations. For energies $Q$ much above the
sphaleron  mass we have found an approximate but explicit solution which 
allowed us to obtain the field configuration at the turning (escape)
time t=0. The approximate solution turns out to be  just a rescaled  Yang-Mills
sphaleron solution~\cite{OCS}. We have solved the corresponding 
Yang-Mills equations describing its Minkowski explosion at later time, 
which is just a rescaled  version of a Luscher-Schechter solution
discussed in~\cite{OCS}. We have used the asymptotic form of the
gauge field in physical gauge, to discuss the gluon number and spectrum.


Using an analytical continuation
of the singular tails into Euclidean time, we have found that the 
same rescaling allows for the determination of the virtual in-gluon
multiplicities both above and below the sphaleron (antisphaleron) point. Thus, 
the in-out multiplicities at the sphaleron and antisphaleron point, implies the in-out
multiplicities away from this point for semiclassical parton-parton
scattering. The scattering liberates about
$1.1/\alpha_s(\rho)$
prompt gluons by sphaleron (antisphaleron) produced. We recall that 
all the calculations are semiclassical, assuming $\alpha_s(\rho)\ll 1$
or large produced multiplicities.
The larger the longitudinal $Q$ transferred
in the CM frame, the smaller the size of the escaping sphaleron or
antisphaleron and vice versa. The semiclassical  production through singular QCD 
solutions in the semi-hard regime, may even extend to the hard regime 
through smaller size sphalerons.  However, in this case the cross
section is small. On the other hand, for typical instantons in the
instanton vacuum relevant to the `semi-hard' scale and possibly to the QCD
pomeron problem, this number turns out to  be only about 3 gluons per
cluster\footnote{There are of course also instantons with smaller
size, larger action and outgoing multiplicity. However in order to
observe those
the price to pay is exponential in action. 
Recall that electroweak sphaleron releases about 50
W,Z,H, which would be quite spectacular, but unobservable for the same 
reason.}. In QCD there would be also quark production which we will
discuss elsewhere.

This mechanism of coherent multi-gluon production is additional to perturbative BFKL
ladders and/or  virtual gluon materialization in
the color-glass approach~\cite{LR}. The difference between this
mechanism and others is seen e.g. in the fact that the released
semiclassical gluons form thin shells of strong coherent field, which
carry topological features inherited from the topological tunneling.
This  difference is important for many applications, including the
production of quarks as we discuss next.

\vskip 1.25cm
{\bf Acknowledgments}
\\\\
This work was supported in parts by the US-DOE grant
DE-FG-88ER40388. RJ was supported in part by KBN grants 2P03B01917 and
2P03B09622.

\appendix

\section{LS solution}

In this section we give a brief characterization of the LS gauge
configuration. The reader is also referred to \cite{ACTOR} for more
details. For the Euclidean Witten ansatz

\eqn
-e A^a_0 = &&\f{x_a}{r} A_0 \\
-e A^a_i = &&\eps_{ian} \f{x_n}{r^2}(1+\phi_2)\nonumber\\&& +\f{x_a x_i}{r^2} A_1
+\left(\dl_{ai}-\f{x_ax_i}{r^2}\right) \f{\phi_1}{r}
\label{WIT}
\eqnx
The coefficient functions for the LS solutions continued to Euclidean
space are
\eqn
A_0 &=& -4 q_{LS} \gm^2 r t \\
A_1 &=& -4 q_{LS} \gm^2 \f{1+r^2-t^2}{2} \\
\phi_1 &=& -4 q_{LS} \gm^2 r \f{1-r^2-t^2}{2} \\
\phi_2 &=& -4 q_{LS} \gm^2 r^2-1
\eqnx

\eq
\gm^2=\f{1}{(1-r^2-t^2)^2+4r^2}
\eqx
and $q_{LS}$ is a function of $\arctanh\left( \f{2t}{1+r^2+t^2}
\right)$. 

In terms of these quantities the electric and magnetic fields are
given by formula\footnote{In the original formula (7.27) the
coefficient of $F_{01}$ is off by a factor of $r^2$.} (7.27) in \cite{ACTOR}.
We perform the calculations for the sphaleron point which corresponds
to
\eq
q_{LS}=-1+\f{\sqrt{2}}{\cos \left( \sqrt{2} \arctanh\left( \f{2t}{1+r^2+t^2}
\right) \right)}
\eqx 

We rotate these formulas back to Minkowski space using $t \to i
t$. The solutions are again concentrated on the light cone. We perform
then the limit $t \to \infty$ keeping $v=r-t$ fixed. The results for
the `electric' coefficients of formula (7.27) in \cite{ACTOR} are
\eqn
D_0 \phi_2/r &\lra& \f{v+2\sech^2 u (\sinh u-v)}{t (1+v^2)^2} \\
F_{01} &\lra& \f{1}{t} \cdot 0 \\
D_0 \phi_1 /r &\lra& \f{1-2\sech^2 u (v\sinh u+1)}{t (1+v^2)^2}
\eqnx
where $u \equiv \sqrt{2} \arctanh(1/v)$. 
The `magnetic' coefficients are
\eqn
-D_1\phi_1/r &\lra& i\f{1-2\sech^2 u (v\sinh u+1)}{t (1+v^2)^2}\\
(1-\phi_1^2-\phi_2^2)/r^2 &\lra&  \f{1}{t} \cdot 0 \\
D_1\phi_2/r &\lra& -i\f{v+2\sech^2 u (\sinh u-v)}{t (1+v^2)^2}
\eqnx
For large times $t$, the electric and magnetic fields are equal

\be
{\vec E}^2 (t,v) \sim {\vec B}^2 (t,v) \sim \frac 2{t^2} \,\frac{1}{(1+v^2)^3}\,\,
\ee
and sums up to $M_S$. The same result was obtained in~\cite{OCS}.

The `antisphaleron' follows from the sphaleron by substituting
\eqn
A_0 &\lra& -A_0 \\
A_1 &\lra& -A_1 \\
\phi_1 &\lra& -\phi_1 \\
\phi_2 &\lra& \phi_2
\eqnx
in the Witten ansatz above, and yields a solution of the YM equation of motion. The same
transformation makes a switch between the instanton and antiinstanton
in the Witten ansatz.

\section{$\im S$ for LS Solution}

In this appendix we will show that the imaginary part of the classical
action along the deformed contour~\cite{PAST} does not depend on the energy of the
LS solution. In other words, the instanton-antiinstanton suppression
factor persits all the way to the sphaleron point in the deformed
contour approach suggested in~\cite{PAST}. Indeed, 
for $\eps<1/2$ the solution is given by the function
\eq
\label{e.ls}
q=q_- \sn (q_+ \phi+K,k)
\eqx 
where we use the same notation as in~\cite{PAST}
\eq
q_\pm = \sqrt{1\pm \sqrt{2\eps}} \qqqq k^2=\f{q_-^2}{q_+^2}
\eqx
and
\eq
\phi= {\rm arccoth} \left( \f{1+r^2+\tau^2}{2\tau} \right)
\eqx
The suppression factor is given by
\eq
\im S= -\f{24\pi^2}{g^2} \int_0^\infty dr \,\im \sum_{nm} \res \left\{
\f{\cos^4 w}{r^2} \EE 
\right\}
\eqx
where
\eq
\EE \equiv  \f{1}{2} \dot{q}^2+\f{1}{2} (q^2-1)^2
\eqx
The residues are calculated at the poles of (\ref{e.ls}) inside
the contour. 
One can show that $\EE$ is just equal to $(q^2-1)^2-\eps$. We will now use the
Laurent expansion of the elliptic sine
\eq
\sn(u) \sim \f{\al}{u-u_0} +\bt (u-u_0)
\eqx 
Using the differential equation for $\sn(u)$ we obtain the relations
$\al^2 k^2=1$ and $\bt=\f{\al}{6} (1+k^2)$. 
We thus have to compute
\eqn
&&\res \left(\cos^4 w \EE \right) =\al^4 k^4 \res \left( \f{\cos^4
w}{(\phi-\phi_0)^4} \right) + \nonumber\\
&&(4\al^3 \bt k^2 q_-^2-2 \al^2 k^2) \res \left( \f{\cos^4
w}{(\phi-\phi_0)^2} \right)
\eqnx
The coefficients of the residues are energy independent and
equal to $1$ and  $-2/3$ respectively. Once we use
\eq
\cos^2 w \equiv \f{4r^2}{(1+r^2)^2 +2\tau^2 (r^2-1)+\tau^4}
\eqx
and the $r$ dependence of the poles $\tau(r)=q+\sqrt{q^2-1-r^2}$ where
$q$ is an $\eps$ dependent constant, we obtain 
\eqn
\int_0^\infty \f{dr}{r^2}\, \res \left(\f{\cos^4 w}{(\phi-\phi_0)^4}
\right)&=&\f{i}{3} \f{1+q^2}{1-q^2} \\
\int_0^\infty \f{dr}{r^2}\, \res \left(\f{\cos^4 w}{(\phi-\phi_0)^2}
\right)&=& i \f{1}{1-q^2}
\eqnx
Finally we get 
\eq
\im S=\f{8 \pi^2}{g^2} 
\eqx
Note that the whole energy dependence through $\eps$ has cancelled out.

\section{Stability under Perturbative Pair Production}

Are the escaping Yang-Mills fields stable under pair production, at
least perturbatively? In general, they include electric as well as
magnetic fields: but the answer can still be given
 by assessing the gluon/quark
pair emission rate in a general time dependent background to first
order in $\alpha$. Let ${\cal T} (A)$ be the on-shell matrix 
element between a particle and an antiparticle in the 
external gauge field. Typically, the number of pairs per 
unit 4-volume is 

\be
\frac{dN_{\overline{q}q}}{d^4x} ={\rm Tr}\la x|\, {\rm ln}\,
\left({\bf 1} -{\cal T} (A)\,{\bf P}(+)\,{\overline{\cal T}} (A)\,{\bf P}
(-)\right) | x \ra 
\label{42}
\ee
where ${\bf P}(\pm)$ are the on-shell
projectors on particles and antiparticles, and the trace
is over color and spin. In the momentum representation

\be
{\bf P} (\pm ) = 2\pi\,(\rlap/{P} + m_F) \,\theta (\pm P_0)\,
\delta(P^2-m_F^2)\,\,,
\label{43}
\ee
for a flavor of mass $m_F$.
Since the pair creation follows from large times (weak fields),
we may approximate ${\cal T} (A)$ by its leading order

\be
{\cal T} (A) \approx ig\rlap/{A}\,\,.
\label{44}
\ee
Inserting (\ref{44}) into (\ref{42}), expanding the logarithm,
integrating over all spacetime, and insisting on gauge invariance 
yield the pair rate

\be
\frac {dN_{\overline{q}q}}{d^4q} =&&\frac{{\bf 4}\alpha}{12\,(2\pi)^4}\,
\,\left(|{\bf E}_i^a (q)|^2 - |{\bf B}_i^a (q)|^2 \right)\nonumber\\
&&\times \theta(q^2-4m_F^2)
\,(1-4m_F^2/q^2)^{1/2}(1+2m_F^2/q^2)\,\,.\nonumber\\
\label{45}
\ee
where $({\bf E}\,,{\bf B})$ are the Fourier transform
of the chromo-electric and -magnetic fields in Minkowski space

\be
({\bf E} , {\bf B} )\, (q)  = \int d^4x\,e^{iq\cdot x}\,
({\bf E} , {\bf B} )\, (x) \,\,.
\label{46}
\ee
We have highlighted ${\bf 4}$ in (\ref{45}) to show 

\be
{\bf 4} = \frac 12 \times 2\times 2\times 2
\ee
for $g/\sqrt{2}$ fundamental quark charge, 
2 flavors, 2 spins and 2 particle-antiparticle
respectively.
Notice that the production in (\ref{45}) is time like, for
which $q = (\omega , \vec{0})$ is an allowed frame. Such 
frames support zero magnetic fields, indicating that the
pair production mechanism is electric in nature. It is of order
$\alpha^0$ in the strong coupling constant. In deriving (\ref{46})
we have ignored the back-reaction of the quarks on the YM fields.

Since the $\overline{u}u$ and $\overline{d}d$ pairs are light,
we may set $m_F\approx 0$ in (\ref{45}). The total
number of light quark pairs emitted by the classical field is

\be
N_{\overline{u}u +\overline{d}d} =&&\frac{\alpha}{3 (2\pi)^4}\,
\int d^4q
\,\theta (q^2)\,\left(|{\bf E}_i^a (q)|^2 - |{\bf B}_i^a (q)|^2
\right)\,\,.
\nonumber\\
\label{x46}
\ee
The present arguments can also be used to derive
a similar expression for the number of gluon pairs emitted by
the classical field in the weak field limit. For $N_c=3$, the
gluons can be organized in 3 conjugate pairs, such as: 
$W^+W^-$, $K^+K^-$, $K^0\overline{K^0}$ by analogy with the
charged octet mesons. Say we choose the external field to be
$K^0$, then $K^0$ may decay into the two charged modes
$W^+W^-$ and $K^+K^-$. Using the background field method, the 
result for the two gluon multiplicity is

\be
N_{gg} =&&\frac{{\bf 3}\,\alpha}{12 (2\pi)^4}\,
\int d^4q
\,\theta (q^2)\,\left(|{\bf E}_i^a (q)|^2 - |{\bf B}_i^a (q)|^2
\right)\,\,,
\nonumber\\
\label{x47}
\ee
We have highlighted ${\bf 3}$,

\be
{\bf 3}= \frac 34 \times 2\times 2
\ee
for $g\sqrt{3}/2$ charge, 2 decay modes and 2 physical
helicities respectively. Combining (\ref{x46}) with (\ref{x47}) 
we find that the light quark and gluon multiplicities are related

\be
N_{\overline{u}u +\overline{d}d}=\frac {2N_F}3 N_{gg} \,\,.
\label{x49}
\ee

Using the results of appendix B, we find

\be
\int d^4q
\,\theta (q^2)\,\left(|{\bf E}_i^a (q)|^2 - |{\bf B}_i^a (q)|^2
\right) =0\,\,.
\ee
The expanding sphaleron starts magnetic and remains magnetic-like
throughout. It is stable under light-pairs quantum emission.


\end{document}